# Effect of epitaxial strain on ferroelectric polarization in multiferroic BiFeO₃ films

Dae Ho Kim[a)] and Ho Nyung Lee
*Materials Science and Technology Division, Oak Ridge National Laboratory, Oak Ridge, Tennessee 37831*

Michael D. Biegalski
*Center for Nanophase Materials Sciences, Oak Ridge National Laboratory, Oak Ridge, Tennessee 37831*

Hans M. Christen
*Materials Science and Technology Division and Center for Nanophase Materials Sciences, Oak Ridge National Laboratory, Oak Ridge, Tennessee 37831*

Multiferroic BiFeO₃ epitaxial films with thickness ranging from 40 nm to 960 nm were grown by pulsed laser deposition on SrTiO₃ (001) substrates with SrRuO₃ bottom electrodes. X-ray characterization shows that the structure evolves from angularly-distorted tetragonal with $c/a \approx 1.04$ to more bulk-like distorted rhombohedral ($c/a \approx 1.01$) as the strain relaxes with increasing thickness. Despite this significant structural evolution, the ferroelectric polarization along the body diagonal of the distorted pseudo-cubic unit cells, as calculated from measurements along the normal direction, barely changes.

Epitaxial films of ferroelectrics have been exten-sively investigated in the pursuit of applications in integrated electronic devices and better understanding of their physical properties.[1] Exploring how epitaxial strain – induced in thin films due to the different lattice constants and/or crystallographic symmetries between film and substrate – manipulates the ferroelectric (FE) properties provides useful insights into the physics of ferroelectricity.[2,3] The most drastic effect has been reported in BaTiO₃ films, where strain is seen to increase the remanent polarization ($P_r$) and transition temperature.[4–6] It is also recognized that the strain response of ferroelectricity strongly depends on the individual material and the specific mechanism.[7] For example, a weak strain-dependence found in tetragonal PbZr₀.₂Ti₀.₈O₃.[8,9] BiFeO₃ is known for its unique FE mechanism, driven by lone-pair electrons of the Bi ion, which is compatible with the antiferromagnetic order of the Fe ions, *i.e.*, multiferroicity.[10,11] This motivates us to investigate the interplay between ferro-electric/magnetic properties and epitaxial strain in BiFeO₃ films.

Bulk BiFeO₃ exhibits a rhombohedrally distorted perovskite structure with space group *R3c*, with a polarization $P \approx 100 \ \mu C/cm^2$ along the [111] direction and antiferro-magnetic ordering of Fe³⁺.[12] In the case of a (001)-oriented film on a substrate with cubic symmetry, such as SrTiO₃, the epitaxial strain will change not only the lattice parameters but also the crystallographic symmetry. In epitaxial thin films, due to the relaxation of the epitaxial strain by formation of structural defects with increasing thickness, the structure of BiFeO₃ films changes with varying thickness.[10,13–16] In this letter, we report a drastic structural evolution in BiFeO₃ films on (001) SrTiO₃ substrates: the films' structure changes from one best described as angularly-distorted tetragonal to one that is similar to the bulk rhombohedral structure as the epitaxial strain relaxes with increasing thickness. Magnetic measurements on the films reveal no signs of ferromagnetic impurities but instead are consistent with an antiferromag-netic order. Surprisingly, the significant structural variation results in a limited change of the FE properties. Assuming a bulk-like *P* orientation along the pseudocubic [111] direction, much of the change in the measured $P_r$ (projected along the substrate normal) can be understood as a consequence of a small rotation of the *P* towards the out-of-plane direction with changing $c/a$ ratio.

Epitaxial BiFeO₃ films with thicknesses ranging from 40 to 960 nm were grown using pulsed laser deposition with a KrF excimer laser ($\lambda = 248$ nm).[17] Films were grown at 700 °C in 50 mTorr of oxygen using a sintered polycrystalline BiFeO₃ target with 15 at. % excess Bi to maintain the stoichiometric content of volatile Bi. The films were grown on single-crystalline SrTiO₃ (001) substrates with coherently grown SrRuO₃ (4 nm in thickness) epitaxial bottom electrodes. The SrRuO₃ layers are fully strained and coherently grown.[18,19] High resolution four-circle x-ray diffraction (HR-XRD) ($\theta$-$2\theta$ and $\phi$ scans, data not shown) confirms that the films are grown as single–phase with an epitaxial relationship: BiFeO₃ (001) ∥ SrTiO₃ (001); BiFeO₃ [100] ∥ SrTiO₃ [100]. Note that throughout this letter, the pseudocubic notation is used for BiFeO₃. To perform *P* measurements, 50 nm thick Pt top electrodes were sputtered to form 50 $\mu$m diameter parallel-plate capacitors. *P* was measured in a cryogenic probe station with a ferroelectric tester (aixACCT TF Analyzer). A superconducting quantum interference device magnetometer (Quantum Design MPMS) was used to investigate the magnetic properties of the films.

XRD reciprocal space maps (RSM) around a SrTiO₃ 113 reflection most clearly reveal the changes in the films' crystalline structure as a function of their thickness: In Fig. 1(a), the peak from the 77 nm thick BiFeO₃ film is perfectly aligned on the horizontal axis with that of the substrate, indicating in-plane lattice match. This strain-induced reduction of the in-plane lattice constant (3.905 Å) from the bulk value (3.962 Å) results in an elongated out-of-plane lattice constant, $c = 4.068$ Å. Peaks in $\theta$-$2\theta$ scans for the (011, $0\bar{1}1$) and (112, $\bar{1}\bar{1}2$) sets of reflections are split (data not shown) resulting from an angular distortion of the tetragonal structure, consistent with a tilt towards the [110] direction by 0.4 ±0.05° (smaller than the value of 0.6° for the

---
[a)]Electronic mail: kimdh@ornl.gov

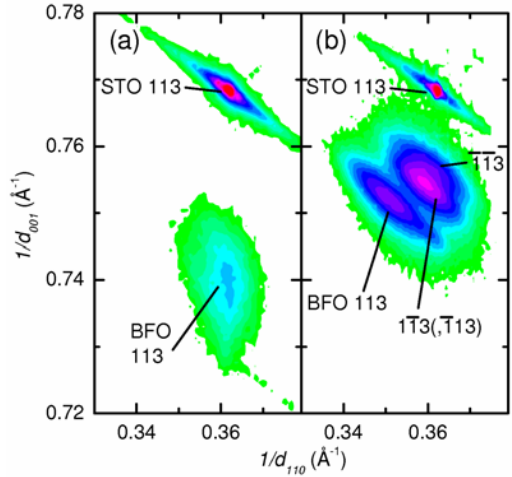

FIG. 1. (Color online) HR-XRD reciprocal space maps around the 113 reflections of (a) 77 nm and (b) 960 nm-thick BiFeO$_3$ films on (001) SrRuO$_3$/SrTiO$_3$ substrates. The vertical and horizontal axes correspond to the out-of-plane and in-plane directions, respectively.

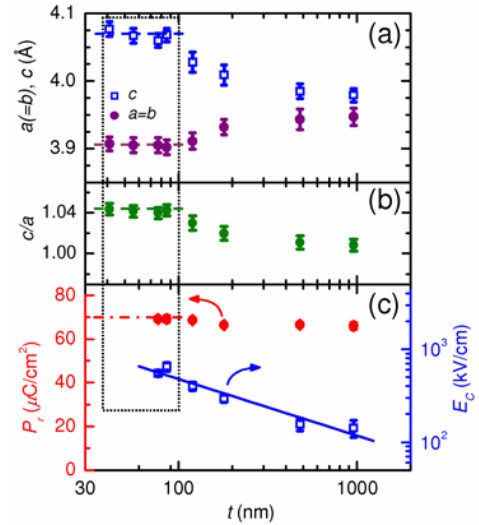

FIG. 2. (Color online) (a) In-plane ($a=b$, purple circles) and out-of-plane ($c$, blue squares) lattice parameters, (b) $c/a$ ratio, and (c) $P_r$ (red circles) and $E_c$ (blue squares) of the BiFeO$_3$ films with varying thickness. The dotted rectangle represents the fully-strained thickness range. The dash-dotted horizontal lines are guides to the eye.

rhombohedral distortion in the bulk). While determining the details of the crystal structure is not necessary for the purpose of this paper, our data for the structure of the 77 nm thick film are thus consistent with a monoclinic (with BiFeO$_3$ [010]$_{monoclinic}$ ∥ SrTiO$_3$ [110]) or lower-symmetry unit cell (Neumann's principle[20] in combination with the previous observation of a polarization direction along the pseudocubic body diagonal[11-13] argues in favor of a distortion along the [110] direction). The observed crystalline structure changes strongly with increasing thickness, reaching a drastically different strain state for a 960 nm-thick BiFeO$_3$ film. The HR-XRD RSM [Fig. 1(b)] shows a splitting of the BiFeO$_3$ peaks into the three values of $2\theta$ for the 113, 1$\bar{1}$3 (equivalent to $\bar{1}$13), and $\bar{1}\bar{1}$3 reflections as would be expected for a rhombohedral structure. The splitting of the 1$\bar{1}$3 and $\bar{1}\bar{1}$3 peaks is weak and barely resolved in our experiments. Further analysis shows that the structure deviates from rhombohedral, with $a = b = 3.947$ Å but $c = 3.979$ Å. Nevertheless, for the purpose of estimating the degree of strain in the film, it suffices to approximate the presumably monoclinic structure as a distorted rhombohedral unit cell with an angular distortion of $0.5 \pm 0.05°$. When compared to the thinner films, this 960 nm-thick film thus exhibits a structure much more similar to that of bulk BiFeO$_3$. These observations of the structural evolution of BiFeO$_3$ films with varying thickness on (001) SrTiO$_3$ substrates confirm and further clarify previously published results on PLD grown films (e.g., the shift of in-plane film peaks' positions in normal *θ-2θ* scans[10,13,14] and samples obtained by MOCVD.[15]

Figure 2(a) shows the dependence of the lattice parameters on the thickness for a series of eight samples. The in-plane and out-of-plane lattice parameters of the films remain unchanged up to a thickness of about 90 nm, as indicated by a dotted rectangle. This observed critical thickness (about 90 nm) is large for a system with such a strong lattice mismatch between film and substrate (~ 1.5 %), even larger than observed in the case of tetragonal PbZr$_{0.2}$Ti$_{0.8}$O$_3$ on SrTiO$_3$ (where a similar mismatch is accompanied by a critical thickness of 40 nm[8]). Possible explanations of this large structural compliance include the system's freedom to undergo tilt transitions. Beyond this critical thickness, the lattice parameters gradually evolve towards the relaxed structure as discussed above. Based on the data in Fig. 2(a), it is found that the $c/a$ ratio decreases from 1.04 (strained) to 1.01 (relaxed) as shown in Fig. 2(b).

Before studying the ferroelectricity of these films, we examine how the magnetic properties depend on the structural variation. The magnetic moment ($m$) vs. magnetic field ($H$) curves of 120 and 960 nm-thick BiFeO$_3$ films measured at room temperature show linear diamagnetic responses from the entire sample, *i.e.*, film and substrate [Fig. 3(a)]. The SrRuO$_3$ bottom electrode is too thin to contribute measurably to the magnetic moment, as discussed elsewhere.[21] The diamagnetic background from the SrTiO$_3$ substrate was calculated and subtracted from the raw data. The resulting $m$ originates from the BiFeO$_3$ film and is shown in Fig. 3(b) as normalized magnetization ($M$) per formula unit vs. $H$ [$M(H)$] data. The corrected data of both films show very small field-induced magnetizations, and the $M(H)$ traces are devoid of any specific features. This is consistent with the antiferromagnetic order observed in bulk BiFeO$_3$ and further indicates the absence of ferromagnetic impurities which could form, for example, as consequence of Bi loss.[22,23]

Turning our attention now to the FE properties, the structural evolution with $c/a$ ratios ranging from 1.01 to 1.04 results only in a very limited change in $P_r$. Figure 3(c) shows $P$ vs electric field ($E$) [$P(E)$] hysteresis loops of 77 nm (strained) and 960 nm-thick (strain-relaxed) BiFeO$_3$ films measured at 77.3 K and 2 kHz. The low-temperature measurements with both films reveal a very similar value of $P_r$ with fully saturated, well-defined square hysteresis loops without appreciable indication of leakage. (Note that measurements at room temperature were not possible due to the electrical conductivity of the films.) The weak dependence of $P_r$ on frequency [Fig. 3(d)] and its saturation with increasing electric field [the inset in Fig. 3(d)] confirm the reliability of these measurements. As expected, the coercive field $E_c$ observed on these $P(E)$ loops is larger for the thinner films. Indeed, $E_c$ exhibits a systematic increase with decreasing the thickness, following a dependence of $E_c$ $\propto d^n$, with n = -0.61 ±0.08 [Fig. 2(c)]. Consistent with previous reports,[8,24] this is in good agreement with the Kay-Dunn scaling law [n = -2/3][25] over the entire film thickness range (77 – 960 nm), and surprisingly unaffected by the change in crystalline structure.

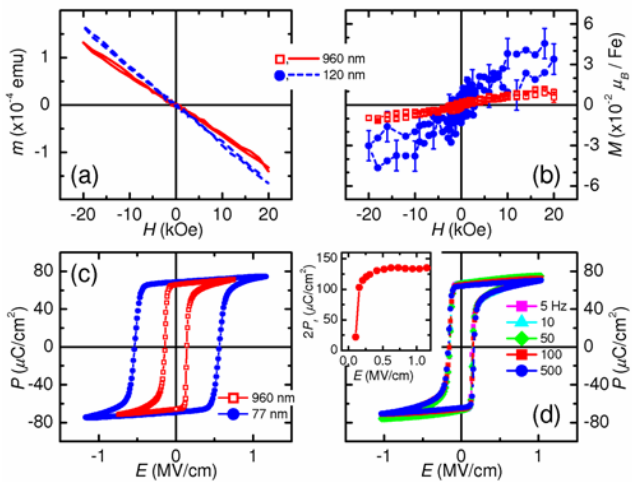

Fig. 3. (Color online) (a) $m(H)$ curves of samples comprised of the BiFeO$_3$ films (120 nm and 960 nm in thickness) and the SrRuO$_3$/SrTiO$_3$ (001) substrates measured at room temperature. (b) $M(H)$ data of the same BiFeO$_3$ films after subtracting the diamagnetic contribution from the substrate. (c) $P(E)$ loops of 77 and 960 nm-thick BiFeO$_3$ films measured at a frequency of 2 kHz. (d) Frequency dependence of $P(E)$ loops of the 960 nm-thick film. The inset in (d) shows $2P_r$ of the same film as a function of the maximum applied $E$. All the $P$ measurements were performed at 77.3 K.

The limited change of $P_r$ with film thickness reveals that the ferroelectricity in BiFeO$_3$ films is almost independent on the specific epitaxial strain induced by (001)-oriented substrates. As shown in Fig. 2(c), the fully strained films (*i.e.*, thinner than 90 nm) exhibit a $P_r$ of 69.2 ±1 $\mu C/cm^2$, whereas for the relaxed film (960 nm), we obtain $P_r$ = 66.0 ±1 $\mu C/cm^2$. $P_r$ values of the films are taken from pulse measurements to eliminate any leakage contributions.[8] Considering the change in structure and making the assumption that the films' $P$ points along the [111] direction as in the bulk, most of this epitaxial "enhancement" of the polarization's projection by 4.8% along the normal direction can be explained by geometric arguments: For our 960 nm-thick film, the measured $P$ along the [001] direction ($P_{001}$) can be seen as a projection of $P_{111} = 115.0$ ±1 $\mu C/cm^2$, whereas for the strained films, this value increases to $P_{111} = 116.9$ ±1 $\mu C/cm^2$. In other words, the higher $c/a$ ratio (and the smaller tilt distortion) in the strained films result in a slight rotation of the $P$ towards the [001] direction, and strain increases the $P_{111}$ by a small value of only 1.6% beyond the geometric effect. These findings agree qualitatively with the previously reported results of first-principle calculations, showing that strained thin films and relaxed thick films on (001) SrTiO$_3$ substrates are expected to exhibit $P_r$ of 63.4 and 57.0 $\mu C/cm^2$, respectively.[16] Comparing these predictions with the presented experimental results and our previous report on robust ferroelectricity in BiFe$_{0.5}$Cr$_{0.5}$O$_3$ films[21] we find that $P_r$ in actual thin-film samples is even less dependent on strain than expected from first-principles calculations, perhaps due to the contribution from other mechanisms (e.g., defects, symmetry changes, etc.) not accounted for in the calculations. The insignificant change in polarization by the experimentally observed, significant structural distortions illustrates the difference between the lone-pair driven ferroelectricity and mechanisms in other perovskite ferroelectrics such as BaTiO$_3$, where the epitaxial strain is known to strongly couple to the ferroelectric polarization. With the very large $P_r$ in BiFeO$_3$, the remarkable independence of the FE properties on strain provides an excellent starting-point for further development of multiferroic systems.

The authors thank P. H. Fleming for technical assistance. Research sponsored by the Division of Materials Sciences and Engineering (DHK, HNL, HMC) and the Division of Scientific User Facilities (MDB), Basic Energy Sciences, U.S. Department of Energy.